\documentclass[11pt]{article}

\input{epsf}
\usepackage{graphicx}
\usepackage{epsfig}
\usepackage{amssymb}

\usepackage{amsfonts}
\usepackage{bbm}

\newcommand{\be}{\begin{equation}}
\newcommand{\ee}{\end{equation}}
\newcommand{\rf}[1]{(\ref{eq:#1})}
\newcommand{\X}{\bar{X}}

\newcommand{\G}{{\cal G}}

\begin{document}

\title{The  Deformable  Universe}
\author{M. D. Maia\thanks{maia@unb.br} and A. J. S. Capistrano\thanks{capistranoaj@unb.br}\\
Universidade de Brasilia, Instituto de Fisica, 70919-970,  Brasilia,  DF.\\
J. S. Alcaniz\thanks{alcaniz@on.br},\\Observat\'orio Nacional, 20921-400, Rio de Janeiro, RJ.\\
Edmundo M.  Monte\thanks{edmundo@fisica.ufpb.br}\\U.F. Paraiba, Departamento de Fisica,  58059-970, Jo\~ao Pessoa, PB.}

\date{\today}

\maketitle

\begin{abstract}

  The concept of   smooth   deformations   of a  Riemannian manifolds,  recently  evidenced  by  the solution of  the Poincaré\'e  conjecture,  is applied  to Einstein's   gravitational  theory  and   in particular  to   the  standard  FLRW cosmology.  We present  a  brief   review   of    the deformation of  Riemannian geometry, showing  how  such    deformations  can be   derived  from  the  Einstein-Hilbert   dynamical   principle. We    show  that  such  deformations  of    space-times  of    general  relativity  produce   observable  effects that  can  be   measured  by  four-dimensional  observers. In the  case of  the FLRW   cosmology,   one  such    observable  effect  is shown  to  be   consistent    with the  accelerated  expansion of  the  universe.

\end{abstract}

\textbf{keywords}:{Geometry,Cosmology,Dark Energy,Geometric Flows}

\tableofcontents

\section{Introduction}
The   $\Lambda$CDM  paradigm    for  the   accelerated  expansion of  the  universe  makes  use    the   cosmological  constant  $\Lambda$, interpreted  as   the vacuum  energy  density   of quantum  fields,    as  the   main  cause  of  the acceleration.
However,  it has  been  proven  to  be very   difficult  to explain
 the large  difference  between  the  very  small  observed  value   $\Lambda/8\pi G \approx  10^{-47} Gev^2/c^4$ and    the  very  large averaged  value of  the  quantum vacuum  energy  density  $<\rho_v> \approx  10^{75}  Gev^2/c^4$.  The    lack of  a  feasible  explanation  for  such   cosmological  constant problem  makes  the  $\Lambda$CDM paradigm unacceptable  as  a preferred   theoretical option.  In face of this  difficulty a    variety  of  alternative explanations  have been  proposed, including   the possible  existence of new and  previously  unheard of  essences;  the  postulation of   specific   scalar   fields;  or  even  the  possible existence  of   non  observable    extra  dimensions in space.

The extra dimensional  proposition is  interesting   because  it may  solve    another  fundamental issue,   namely  the  hierarchy of the fundamental  interactions, the   huge ratio of  the   Planck to   the electroweak energy  scale ($M_{Pl}/M_{EW} \sim 10^{16}$).   Indeed,   Newton's  gravitational  constant $G$   depends on the dimension of  space.  It  has been     shown   that     in  a    higher  dimensional  space   the    constant  $G$   must  change   to  another  value $G_*$,  such that  gravitating  masses   can be correctly   evaluated  by a (higher  dimensional) volume integration of   given  mass  densities \cite{ADD}.

Yet,   the hypothetical  existence  of  extra  dimensions must be compatible  with  the  experimentally proven  and  mathematically  consistent  four-dimensionality  of   space-times.
 For  example it took  about   60   years to  find  out  that   the  Kaluza-Klein theory based on the  Einstein-Hilbert  principle and having a  product  topology  space,     is   not    compatible  with  the observed  fermion  chirality  at the  electroweak  scale,   mainly   because  the   diameter  of  the  compact internal  space  is  too  small  (the Planck  length).

 In  a  more recent  proposal     the   product  topology  of  the  higher  dimensional  space   has  been   replaced  by  an  embedding  space,  while  the  Einstein-Hilbert principle  was  maintained  and  some  other   assumptions  are introduced.  The  four-dimensionality  of  space-time  is  maintained,  but    the  gravitational field   propagates not  only  in the  four-dimensional   space-time  but  also  along   the  extra  dimensions.
However  the  dynamics of  this  extra-dimensional      propagation  or  deformation,    has not  been  detailed   and  this  is   the   main  subject of   this  paper.

Several   interesting  models  have  been   proposed,  mostly  belonging  to  the  brane-world  paradigm  proposed in  \cite{ADD,Rubakov},  sometimes  using additional  conditions \cite{RS,RS1},  or  other  specific   embedding  assumptions  as  for  example  in
 e.g.~\cite{DGP,Sahni,BW1,BW2,BW3,BW4,BW5,BW6,BW7,BW8,BW9}.
In  spite  of  such efforts we  still do  not have a   model  independent  solution of  the  present  cosmological  problems \cite{Goenner}.

The  purpose  of  this    paper  is  to   study    the  dynamics  of  deformation  of  gravitational  fields  in  arbitrary  directions.    We  will see that  such  deformations  are    associated  with  a  conserved  quantity,    the   deformation tensor,   which leads  to  an  observable  effect in  space-time.   We  will  show   that     the  current  observations  on the acceleration of  the  universe   are  consistent  with  the  observational  effect   of  the  deformation  tensor.

\section{Smooth  Deformations of Space-times}

The  concept of  smooth   deformation  of   Riemannian  manifolds  was    defined   by   John  Nash as   a   means  to  correct  the  inability   of   the Riemann tensor  to  distinguish  the  local  shape  of  the  manifold.   This    problem   lies  at the foundations  of  Riemannian  geometry  and  it is  worth  reviewing  it,  starting  from  Riemann's  own  words   as  we  quote: \emph{...We  may,  however,  abstract  from  external  relations  by considering  deformations   which  leave  the  lengths of  lines  within  the  surfaces unaltered, i. e,  by  considering    arbitrary   bendings -without  stretching- of  such  surfaces,  and  by  regarding all surfaces obtained  from one  another in this  way    as  equivalent. Thus, for  example,  arbitrary  cylindrical  or  conical  surfaces  count as  equivalent  to a  plane...} B.  Riemann \cite{Riemann}.

In  the  application of  Riemannian  geometry to Einstein's  gravitational  theory,  the   observables  of  the  gravitational  field are   determined by  the  eigenvalues  of  the  Riemann  tensor  (or its  trace-free  Weyl  tensor  for  pure gravitation),   with  respect  to  the zero  gravitational  field  of  the flat plane  Minkowski  space-time  of  special  relativity.  However as  pointed out  by  Riemann,  the  same  tensor also vanishes   for   cones, ruled hyperboloids,   or  for   helicoidal  space-times. This  leads  to the  conclusion that    in general  relativity  the differences  between  these  shapes are  not  relevant  to gravitation (see   e.g. \cite{Wheeler}).   We  will  show that they   can actually  be  detected   by  an  observer in  a  four-dimensional  space-time.

A  general  solution for the  shape   problem   in  Riemannian  geometry  was  suggested    by   L. Schlaefli in  1871,  proposing    that  \emph{all Riemannian  manifolds must be    embedded in a larger  space},  in  such  a  way  that their   Riemann tensors   would  be   compared  with  the geometry  of  the  embedding  space. Specifically,  the  local  shape of  a Riemannian  manifold  is obtained by  the difference
between   the  Riemann   tensors  of  the  embedded and  the  embedding manifolds (in the  original  proposition  the  embedding  space  was  assumed  to be  flat) \cite{Schlaefli}.  Most  importantly,   Riemannian  geometry  is  recovered  with the  application of  the  inverse  embedding  map.

however, such     solution of the  shape  problem in Riemannian  geometry  depends   on    solving    the Gauss-Codazzi-Ricci  equations,  which  are    non-linear  differential equations  involving    the  metric, the extrinsic  curvature and  the  third fundamental  form  as  independent  variables.  They  provide  the necessary and  sufficient conditions  for  the  existence  of  the embedding   functions  for  a  given  Riemannian  manifold \cite{Eisenhart}.
Until very   recently   only   particular   solutions  of  those  equations were  obtained  with    the help   of    positive power  series  expansions of the embedding  functions differentiable or  by  try and  error.

Nash's  theorem of  1956 changed  this  picture  when he  proposed that the  metric  of   a  given  Riemannian  manifold    could  be  smoothly  deformed  along  an  orthogonal  directions  with  parameter  $y$,  given by
\be
k_{\mu\nu} =-\frac{1}{2}\frac{\partial  g_{\mu\nu}}{\partial y}\label{eq:geometricflow}
\ee
where   $k_{\mu\nu}$  denotes  the  extrinsic  curvature   and
$y$ represents   a    coordinate on a  direction   \emph{orthogonal to  the   embedded}  geometry \cite{Nash}.  Thus,  Nash's
theorem   introduced   the  concept of   deformable  Riemannian  manifolds  in arbitrary  directions,  at the  same  time  that  it
 solved  the  embedding problem.

The  condition  \rf{geometricflow}   is  a  generalization  of  the  well  known  York relation used in the  study of  the  initial  value problem  for  3-dimensional  surfaces in  general  relativity  \cite{York},  to  the case  where  $y$ is not   necessarily  the  time coordinate. It  is  also     analogous,  but  far  more general  than  the `` Ricci flow"  condition   proposed  much  latter  by  R.  Hamilton  using  the    Fourier heat  flux law to  obtain  the  expression \cite{RHamilton}
 $$
 R_{\mu\nu} =-\frac{1}{2}\frac{\partial  g_{\mu\nu}}{\partial y}
 $$
where  $y$  represents  any  coordinate  of a  3-dimensional   manifold.
This  result  was  subsequently  applied  with   success   by   G. Perelman  to  solve  the  Poincaré\'e  conjecture \cite{Perelman}.  Unfortunately  this    condition is  not  relativistic  and  it is  not  compatible  with  Einstein's  equations or  with  relativistic   cosmology.   Indeed,  together  with  Einstein's equations,  the  above   equations  gives  a linear  equation  for  the  gravitational field  with respect  to  an  arbitrary space-time  direction  $y$, strongly   constraining   the  propagation of  gravitation  to
\[
\frac{\partial g_{\mu\nu}}{\partial  y}=  -16\pi G (T_{\mu\nu}-\frac{1}{2}T  g_{\mu\nu})
\]

On the other  hand,     \rf{geometricflow}   does not  have such  limitation because  in  each  embedded   space-time   $g_{\mu\nu}$  and  $k_{\mu\nu}$   are independent variables  satisfying  the Gauss-Codazzi-Ricci  equations,  instead of  \rf{Ricciflow}.   In the  following  we present a    derivation  of   \rf{geometricflow}   for    the  simple  case of   just one  extra  dimension.  Higher  dimensional  cases    were  also  implicit  in  Nash's  paper,  and  it  was  applied  as  a possible
extension of  the  ADM   quantization of  the  gravitational  field \cite{QBW}.

Consider  a  Riemannian  manifold  $\bar{V_n}$ with  metric $\bar{g}_{\mu\nu}$,  and its  local  isometric  embedding in a   D-dimensional  Riemannian  manifold   $V_D$,  $D= n+ 1$, given  by  a   differentiable  and  regular map $\bar{X}: \bar{V}_n \rightarrow  V_D$  satisfying the  embedding  equations\footnote{Throughout  the  paper, except   when  explicitly  stated in  contrary ,  we  will   use  $D=5$  with  metric  signature  $4+1$.  Capital  Latin  indices  run  from 1  to  5 and four  dimensional   indices are  denoted by  Greek  letters. }
\be
\X^A{}_{,\mu} \X^B{}_{,\nu}\G_{AB}=g_{\mu\nu},\;  \X^A{}_{,\mu}\bar{\eta}^B \G_{AB}=0,  \;  \bar{\eta}^A \bar{\eta}^B \label{eq:X}\G_{AB}=1,\;
A,B = 1..D
\ee
 where  we have  denoted by   $\G_{AB}$  the metric components of  $V_D$  in  arbitrary  coordinates,  and where  $\bar{\eta}$  denotes  the unit  vector  field   orthogonal  to  $\bar{V}_n$. The   extrinsic  curvature of  $\bar{V}_n$  is by  definition  the   projection of  the  variation of $\eta$   on the tangent plane \cite{Eisenhart}:
\be
\bar{k}{}_{\mu\nu} =  -\bar{X}^A{}_{,\mu}\bar{\eta}^B{}_{,\nu}  \G_{AB}=\X^A{}_{,\mu\nu}\bar{\eta}^B \G_{AB} \label{eq:extrinsic}
\ee
The  integration of the   system of  equations  \rf{X}   gives  the   required   embedding  map  $\bar{X}$.

Next,  construct   the    one-parameter  group of  diffeomorphisms  defined    by the  map   $h_{y}(p): V_D\rightarrow V_D$,  describing  a  continuous curve    $\alpha(y)=h_y (p)$, passing  through  the  point  $p \in \bar{V_n}$,  with  unit    normal  vector  $\alpha'(p) =\eta(p)$.   This    group  is  characterized by   the composition  $h_{y} \circ h_{\pm y'}(p)\stackrel{def}{=} h_{y \pm y'}(p)$,  $h_{0}(p)\stackrel{def}{=}p$.
Applying   this    diffeomorphisms    to   all points of  a
neighborhood of  $p$, with  a  smooth  variation of  the  parameter  $y$  (regardless   if   the parameter  $y$  is  time-like or  not, or  if it  is   positive or  negative),   we obtain     a   congruence   of  curves (the orbits of the group),   all orthogonal  to   $\bar{V}_n$,  describing  a   smooth   flow  of  points in  $V_D$, which  may (or  not) define  the   deformed  manifold  $V_n$.

Given   a  geometric  object $\bar{\omega}$   in  $\bar{V}_n$,  its Lie  transport along that   flow   for  a small distance  $\delta y$    is  given  by     $\Omega   = \bar{\Omega}  + \delta y  \pounds_\eta{\bar{\Omega}}$,  where $\pounds_\eta$  denotes the
   Lie  derivative  with respect  to   $\eta$
\cite{Crampin}.   In particular,   take  the  Lie  transport of the  Gaussian frame    $\{\X^A_\mu ,  \bar{\eta}^A_a  \}$  of the original manifold   $\bar{V}_n$   obtaining
\begin{eqnarray}
Z^A{}_{,\mu}  &=&  X^A{}_{,\mu}  +   \delta y \;\pounds_\eta{X^A{}_{,\mu}}
% =
% X^A{}_{,\mu}
%+ \delta y \;[\eta_{,\mu}, X]^A
=  X^A{}_{,\mu} + \delta y \;  \eta^A{}_{,\mu}\label{eq:pertu1}\\
  \eta^A  &=&\bar{\eta}^A  +   \delta y\;[\bar{\eta}, \bar{\eta}]^A
\;\;\;\;\;\;= \;\;\bar{\eta}^A \label{eq:pertu2}
\end{eqnarray}
However,  it  should  be  noted  from    \rf{extrinsic}   that     in general   $\eta_{,\mu}  \neq \bar{\eta}_{,\mu}$.

The  set of  coordinates  $Z^A$   obtained by   integrating  these   equations  \emph{does not necessarily   describe  another  manifold}.  In order  to be  so,  they need  to  satisfy
embedding  equations similar  to  \rf{X}:
\be
Z^A{}_{,\mu} Z^B{}_{,\nu}\G_{AB}=g_{\mu\nu},\;  Z^A{}_{,\mu}\eta^B \G_{AB}=0,  \;  \eta^A \eta^B \G_{AB}=1 \label{eq:Z}
\ee
 Replacing    \rf{pertu1} and   \rf{pertu2} in  \rf{Z} and  using  the  definition  \rf{extrinsic} we obtain  the    metric  and  extrinsic  curvature of  the   new  manifold
\begin{eqnarray}
&&g_{\mu\nu} =   \bar{g}_{\mu\nu}-2y \bar{k}_{\mu\nu} + y^2 \bar{g}^{\rho\sigma}\bar{k}_{\mu\rho}\bar{k}_{\nu\sigma}\label{eq:g}\\
&&k_{\mu\nu}  =\bar{k}_{\mu\nu}  -2y \bar{g}^{\rho\sigma} \bar{k}_{\mu\rho}\bar{k}_{\nu\sigma}  \label{eq:k1}
\end{eqnarray}
It  is  easy  to  see that Nash's  deformation   condition  \rf{geometricflow}  follows  from the   derivative  of  \rf{g} with  respect  to  $y$  and   comparing the result  with  \rf{k1}.

Of  course, in order  to  define   a  new differentiable manifold, equations \rf{Z}  need  to be integrated.    The  integrability  conditions  for   these  equations  are  intimately  associated  with the   differentiable (smooth)  properties of  the  embedding  functions,  providing the  proposed  solution of   the
shape problem.  That is,    the   components of  the  Riemann  tensor of  the embedding  space\footnote{To  avoid  confusion  with  the  four dimensional  Riemann tensor $R_{\alpha\beta\gamma\delta}$,  the   five-dimensional Riemann tensor  is  denoted  by    $^5{\cal R}_{ABCD}$. The   extrinsic  curvature terms  in  these equations follows from  the five-dimensional Christoffel  symbols together with the  use of \rf{geometricflow}. },  are  evaluated in the  Gaussian  frame $\{ Z^A_\mu, \eta^A  \}$
\begin{eqnarray}
&&\hspace{-7mm}^5{\cal R}_{ABCD}Z^A{}_{,\alpha}Z^B{}_{,\beta}Z^C{}_{,\gamma}Z^D{}_{,\delta} = R_{\alpha\beta\gamma\delta} +\!\!
(k_{\alpha\gamma}k_{\beta\delta}\!-\!
k_{\alpha\delta}k_{\beta\gamma})\!\!
%\frac{\Lambda_*}{6} (g_{\alpha\gamma}g_{\beta\delta}-g_{\alpha\delta}g_{\beta\gamma})
\label{eq:G1}\\
&&\hspace{-7mm}^5{\cal R}_{ABCD}Z^A{}_{,\alpha} Z^B{}_{,\beta}Z^C{}_{,\gamma}\,\eta^D=k_{\alpha[\beta;\gamma]}  \label{eq:C1}
\end{eqnarray}
We  obtain the      Gauss-Codazzi  equations  \cite{Eisenhart}.  The  first  of these  equation (the Gauss  equation) clearly  shows  that the  Riemann   curvature of  the  embedding  space  acts  as  a  reference  for  the  Riemann  curvature  of  the   embedded  space-time.   It is  true  that both   Riemann  curvature  tensors  carry    the  same   shape   problem     in the  sense  described  by  Riemann,  but   the   differences  between   the two  Riemann  tensors  given  by the extrinsic  curvature  defines  the    shape of the embedded  geometry.
The  second  equation (Codazzi) complements  this  interpretation,  stating  that projection of  the  Riemann  tensor  of  the embedding space  along the  normal  direction  is  given  by  the tangent variation of  the  extrinsic  curvature.
Although the  normal  vector  $\eta$  is  not  observable,    the  extrinsic   curvature  is   quantity  defined in  space-time.   A  third  equation,  the Ricci  equation ,    is  a trivial  identity  in the  case of  just one  extra  dimension  (hypersurfaces).

\section{Deformation  Dynamics}
Equations   \rf{g} and  \rf{k1}  describe  the metric  and  extrinsic  curvature  of  the  deformed   geometry ${V}_4$.   By  varying  $y$  they   can   describe   a  continuous sequence   of  deformed  geometries.    The   existence
of  these deformations are given  by   the  integrability conditions \rf{G1}   and  \rf{C1}.  as  such  these  equations  must  not be  confused   with  dynamical  equations.

   As in Kaluza-Klein   and    brane-world  theories,  the  embedding  space   $V_5$  has  a  metric  geometry  defined   by  the  higher-dimensional Einstein's  equations
\be
^5{\cal R}_{AB} -\frac{1}{2} \,{^5{\cal R}} {\cal G}_{AB}  =G_{*}  T^*_{AB} \label{eq:BE0}
\ee
where  $G_* $  is the  new  gravitational  constant  and  where $T^*_{AB}$ are  components of   the energy-momentum tensor of the  known material sources.  These  equations  are derived from the  Einstein-Hilbert  principle,  to  which  we give  a  natural  interpretation:   the  space-times  satisfying  \rf{BE0}  are  those  with the  smoothest  Riemannian curvature.
\[
\delta  \int{^5{\cal R}} \sqrt{\cal G}  dv =0
\]
From \rf{BE0}     we   may  derive  the gravitational  field  in the embedded  space-times,  after   the  following  observations\vspace{1mm}\\
1)    A  cosmological  constant   was not  included in  \rf{BE0},
so  that   the  existence  of   an  embedded     4-dimensional  Minkowski  space-time (A  cosmological  constant   was  included in \cite{GDEI},  but here  we  see no  reason for  it.).   With this  choice  we  also  ensure that the  cosmological  constant problem does not appear.
\vspace{1mm}\\
2) The  confinement  of  the  gauge  fields to  four  dimensions
is  not   an  assumption, but   a  consequence  of the fact that  only  in  four  dimensions  the  three-form   resulting  from the  derivative of  the Yang-Mills  curvature   tensor  is  isomorphic  to the    one-form current. Consequently,  all  known  observable  sources  of gravitation  composing   $T_{AB}$ are  necessarily    confined to    four-dimensional   embedded  space-times.
Such    confinement   can  be  implemented  a   very  simple   way  by  writing  Einstein's  equation  \rf{BE0}  in  the  Gaussian  frame  of  every   space-time  with the   energy-momentum  tensor   source $T_{\mu\nu}$    is   such  that
\be
8\pi G T_{\mu\nu}  =G_* Z^A_{,\mu}Z^B_{,\nu}T^*_{AB},\;\; \;  Z^A_{,\mu}\eta^B T^*_{AB}=0, \;\;\; \mbox{and}\;\;   \eta^A  \eta^B T^*_{AB}=0  \label{eq:confinement}
\ee
\vspace{1mm}\\
3)  The  set  of  all  deformations  of  a given  space-time   generates  a  continuous   foliation of  the  embedding  space,   composed  by   four-dimensional  space-times,  parameterized  by
the extra  dimension $y$.   For  each  fixed  value of  $y$,  we  obtain  a  deformed   space-time  which, if   so  desired  can  be  de-embedded,  with the  application of  the local inverse  embedding map,   which always  exists   provided  the  embedding is  regular.
In  this  way  we  may  recover  the  purely intrinsic  Riemannian
geometry.  In  some  models   the   addition  of   extra  conditions  may prevent  not  only  the  construction of  the  foliation,  but  also   the  recovery of  the Riemannian   structure.
One particular class   of    models (e.g.  \cite{RS,RS1})  uses  the  Israel-Lanczos  boundary  condition  \cite{Israel}
\be
k_{\mu\nu}=  G_*  (T_{\mu\nu}-\frac{1}{3}Tg_{\mu\nu})  \label{eq:Israel}
\ee
When applying  Nash's  deformation  we  cannot  have  such  condition.  In the first place  because it    fixes  once for  all  the  value  of the  extrinsic  curvature  in terms  of  the  confined  sources,  thus  preventing  the   application of  \rf{geometricflow}.    The  condition  \rf{Israel}  is  also  limited  to  hypersurfaces,  so  that  if  the embedding
 requires  additional  dimensions  it  does not apply.  In  addition  to   obtain  \rf{Israel}   we  also  require   a  special  that  the embedded  space-time  is   a  fixed  boundary   between  two  sides   of   the  embedding  space  with mirror symmetry.  To see this,  consider  Einstein's  equations  in  five  dimensions,  \rf{BE0}  which  can  be written as
\be
^5{\cal R}_{AB}=  G_* (T^*_{AB}  -\frac{1}{3}  T^* {\cal G}_{AB})
\label{eq:BE01}
\ee
 the  left  hand  side  may  be  evaluated  in  the  embedded  space-time  frames  by  contracting it  with  $Z^A_{,\mu} Z^B_{,\nu}$, using   \rf{geometricflow},  \rf{Z}  and the  confinement  conditions  \rf{confinement}, obtaining  the tangent components
 \be
^5{\cal R}_{\mu\nu}=  R_{\mu\nu}  + \frac{\partial  k_{\mu\nu}}{\partial y}  -2k_{\mu\rho}k^{\rho}_{\nu}  +  hh_{\mu\nu} =8\pi G_* (T_{\mu\nu}-\frac{1}{2}T g_{\mu\nu} ) \label{eq:Ricciembedded}\\
\ee
As we can  see,   \rf{Ricciembedded}  does  not  coincide with  \rf{Israel}.   In order to  obtain  the  Israel-Lanczos   condition   from the  above  equations   it  becomes  necessary  to  fix
the embedding, say   at   $y=0$;   find  the  values of \rf{Ricciembedded}   on  both  sides  and  finally
evaluate  the difference  between these  values.  We  find  that   all tangent  components   cancel,  except  the terms   $ \partial k_{\mu\nu}/ \partial y$,   which  add when the   $y$  change  sign  from one  side to another  of the  boundary  $y=0$. Finally,    by   integrating  that   difference  in $y$,  using a  Dirac's  function on  $y=0$,  we  obtain \rf{Israel}. In some   models    motivated  by  string theory,  the  condition \rf{Israel} is  imposed  upfront,  making  it  impossible  to    conciliate  those  models  with    Nash's    deformations.

With  these    remarks  we may   proceed  with the  deformation  dynamics,  now  contracting   \rf{BE0} in its original  form    with  $\{ Z^A_{,\mu},  \eta^A\}$ using \rf{Z}  and  the    confinement conditions  obtaining  two gravitational  equations
(These  are  the same  equations derived   in \cite{GDEI})
\begin{eqnarray}\label{eq:BE1}
&&R_{\mu\nu}-\frac{1}{2}Rg_{\mu\nu}-Q_{\mu\nu}=
8\pi G T_{\mu\nu}\; \hspace{2mm}\\
&&\label{eq:BE2} k_{\mu;\rho}^{\;\rho}-h_{,\mu} =0\;,\hspace{4,9cm}
\end{eqnarray}
where     $h^2= g^{\mu\nu}k_{\mu\nu}$ is  the  (squared) mean curvature and
 $K^{2}=k^{\mu\nu}k_{\mu\nu}$  is  the  (squared)  Gauss curvature
and  where  the  term   $Q_{\mu\nu}$ is
\begin{equation}\label{eq:qmunu}
Q_{\mu\nu}=g^{\rho\sigma}k_{\mu\rho }k_{\nu\sigma}- k_{\mu\nu }h -\frac{1}{2}\left(K^2-h^2\right)g_{\mu\nu}\;,
\end{equation}
This  geometrical  quantity   called   the   deformation  tensor  is  conserved in the sense of
\begin{equation}\label{eq:cons}
Q^{\mu\nu}{}_{;\nu}=0\;.
\end{equation}
This  means  that    there  are   observables effects associated  with the  extrinsic  curvature  in the  four-dimensional  space-time.

To  understand   the  nature of  the   observables  associated  with   the extrinsic  curvature,
consider again  the  one-parameter  group of  diffeomorphism  defined  by  points  in  an  embedded  space-time, and  the  unit  normal  vector   $\eta$,  with orbit  $\alpha(y) = h_{y}(p)$.
 The  Frenet equation  for this  orbit   tells  that there is  a transverse acceleration orthogonal  to its  velocity $\eta$,  which is   therefore  tangent  to  the  embedded  space-time.   As  such,  this vector  can  be  written as  a  a linear  combination of  the tangent basis   $\{  Z^A_{,\mu} \}$  expressed  as
\be
\eta^A_{,\mu}=  g^{\rho\sigma}k_{\mu\rho} Z^A_{,\sigma}  \label{eq:orbitacceleration}
\ee
As  it happens,   except  for  a  difference in  sign   this  is  the  definition  of   the  extrinsic  curvature (see  e.g.  \cite{Eisenhart}.).
\emph{Therefore,  the presence of the  extrinsic  curvature    associated  with  \rf{geometricflow} represents  an  acceleration tangent  to  space-time}.  Since   such  acceleration   always  points  to  the  concave  side  of  the  curve,  then   in  the  case of   a  deformation    with   volume  expansion,    it  implies  in   the   emergence of
the  Riemann   stretching  on  the  space-time geometry,  which in principle  can  responsible  for  the  accelerated expansion of  the  universe.

Nash's   deformation  condition \rf{geometricflow}  tells  how
the  embedding  space  can    be  filled  by   a  continuous  succession of deformed  space-times,   each one  given  by  a fixed value of  $y$. In  each  of  these  space-times  the   metric $g_{\mu\nu}$  and the extrinsic  curvature $k_{\mu\nu}$  are     independent variables   satisfying the  Gauss-Codazzi  equations.   Therefore  each of  them requires the determination of   20  unknowns,  whereas counting  from   \rf{BE0} we  have  only  15  dynamical  equations. If  we ignore  $k_{\mu\nu}$    in   \rf{BE1},  we  obtain  the  usual  Einstein's  equations for  the  metric $g_{\mu\nu}$,    suggesting  that    the   missing   equations  describe  the  extrinsic  curvature.

  Since $k_{\mu\nu}$  is a    symmetric  rank-2  tensor, it  corresponds also  to a  spin-2 field  whose  dynamics is
determined  by a  well  known  theorem due  to   S. Gupta. It   tells   that any  such    tensor  necessarily    satisfy an  Einstein-like  system of  equations, having  the  Pauli-Fierz  equation  as its linear  approximation \cite{Salam,Gupta,Fronsdal}.
 The original  theorem of  Gupta  was  set in the Minkowski  space-time.   Here  we need  to  derive    Gupta's  equations  for    the extrinsic  curvature  in     a  deformed space-time  with  metric $g_{\mu\nu}$.

    Using an   analogy  with the  derivation  of  Einstein's  equations,  we start by  noting that  $k_{\mu\nu}k^{\mu\nu}  =K^2  \neq  4$, so  that  we   need  to normalize   the  extrinsic  curvature, defining  a  temporary  tensor
\be
f_{\mu\nu} = \frac{2}{K}k_{\mu\nu},
\label{eq:fmunu}
\ee
and define  its  inverse by $ f^{\mu\rho}f_{\rho\nu} =   \delta^\mu_\nu$.  It  follows  that    $f^{\mu\nu}=\frac{2}{K}k^{\mu\nu}$.

Denoting  by  $||$   the covariant derivative with respect to   a  connection  defined by $f_{\mu\nu}$,  while  keeping the usual semicolon notation for the covariant derivative with respect to $g_{\mu\nu}$,  the   analogous to the  ``Levi-Civita"  connection associated with $f_{\mu\nu}$  such   that ''  $f_{\mu\nu||\rho}=0$,  is:
\be
\Upsilon_{\mu\nu\sigma}=\;\frac{1}{2}\left(\partial_\mu\; f_{\sigma\nu}+ \partial_\nu\;f_{\sigma\mu} -\partial_\sigma\;f_{\mu\nu}\right)  \label{eq:upsilon}
\ee
Defining
$$
\Upsilon_{\mu\nu}{}^{\lambda}= f^{\lambda\sigma}\;\Upsilon_{\mu\nu\sigma}
$$
The  ``Riemann tensor'' for  $f_{\mu\nu}$ has  components
$$
{\cal  F}_{\nu\alpha\lambda\mu}= \;\partial_{\alpha}\Upsilon_{\mu\lambda\nu}- \;\partial_{\lambda}\Upsilon_{\mu\alpha\nu}+ \Upsilon_{\alpha\sigma\mu}\Upsilon_{\lambda\nu}^{\sigma} -\Upsilon_{\lambda\sigma\mu}\Upsilon_{\alpha\nu}^{\sigma}
$$
and  the  analogous  to the  ``Ricci tensor'' and the ``Ricci scalar'' for  $f_{\mu\nu}$ are, respectively given  by
$$
{\cal  F}_{\mu\nu} =  f^{\alpha\lambda}{\cal  F}_{\nu
\alpha\lambda\mu}
%\;\mathcal{F}_{\mu\lambda\nu}^{\;\;\;\lambda}\;=\mathcal{F}_{\mu\nu}
\;\;\mbox{and}\;\;{\cal  F}=f^{\mu\nu}{\cal  F}_{\mu\nu}
$$
Finally,  Gupta's equations for $f_{\mu\nu}$  can be  obtained  from the  contracted  Bianchi  identity
\begin{equation}
\label{eq:gupta}
{\cal  F}_{\mu\nu}-\frac{1}{2}{\cal  F} f_{\mu\nu}
%+  \Lambda_f f_{\mu\nu}
=\;\alpha_*  \mathcal{T}_{\mu\nu}
\end{equation}
where $\mathcal{T}_{\mu\nu}$ represents  the  source of this field  such that   $\mathcal{T}^{\mu\nu}{}_{||\nu}=0$ and   $\alpha_*$  is  a  coupling  constant.  Notice  that   in  spite  of  the  resemblances,       $k_{\mu\nu}$ is  not  a  metric  because it  exists  only  after  the  Riemannian  geometry  with  the metric  $g_{\mu\nu}$  has  been  defined  for  the  metric  $g_{\mu\nu}$.

\section{Deforming   the FLRW  Universe}

As  we  have seen  Nash's  deformations of  a  space-time
defined  by the  extrinsic  curvature satisfying Gupta's  equation  produces  a   tangent    acceleration in   space-time.   We  have seen  also that the same extrinsic   curvature produces  an  observable  quantity  $Q_{\mu\nu}$.
Such  reasoning   non-trivial  sequence  suggests  that  the currently  observed
acceleration  of the   distant  supernovae   type Ia  (SN Ia), can  be  related  to  the  deformations  of  the standard FLRW  universe,  something    that has  to be     experimentally  verified.

For  that purpose  consider  the  line  element   of the  FLRW   universe   written as
\[
ds^{2}=g_{\alpha\beta}dx^{\alpha}dx^{\beta}=-dt^{2}
+a^{2}[dr^{2}+f(r)(d\theta^{2} +sen^{2}\theta d\varphi^{2})]
\]
where  $f(r)=\sin r, r, \sinh r$   corresponds  to the spatial
curvature  $k=1,0,-1$ respectively.  The
the confined  source  is the  perfect fluid given  in co-moving coordinates   written  as
 \begin{equation}
T_{\alpha\beta}=(p+\rho)U_{\alpha}U_{\beta}
+pg_{\alpha\beta},\;\;U_{\alpha}=\delta_{\alpha}^{4}. \label{eq:T}
\end{equation}
The  embedding   of  the  FLRW   universe  in  a  five  dimensional   flat  space   gives  the  solution  (details  in  \cite{GDEI})
\begin{eqnarray}
&&k_{ij}=\frac{b}{a^2}g_{ij},\;\;i,j=1,2,3, \;\;\;\; k_{44}=\frac{-1}{\dot{a}}\frac{d}{dt}\frac{b}{a}\;, \label{eq:k}
\end{eqnarray}
Just  for  notational  simplicity denote  $b=-k_{11}$, $\xi = k_{44}$,  $H = \dot{a}/a$ and
 $B= \dot{b}/b$.  Then    the  components of the extrinsic  geometry   can  be  written  as
\begin{eqnarray}
\label{eq:BB}
 &&
 %k_{ij}=\frac{b}{a^2}g_{ij},\;\;
 \xi=  \frac{b}{a^{2}}(\frac{B}{H}-1)g_{44},\;\;  \\
&&K^{2}=\frac{b^2}{a^4}\left( \frac{B^2}{H^2}-2\frac BH+4\right),
 \;\;\, h=\frac{b}{a^2}(\frac BH+2)\label{eq:hk}\\
&&Q_{ij}= \frac{b^{2}}{a^{4}}\left( 2\frac{B}{H}-1\right)
g_{ij},\; \quad \quad Q_{44} = -\frac{3b^{2}}{a^{4}},
  \label{eq:Qab}\\
&&Q= -(K^2 -h^2) =\frac{6b^{2}}{a^{4}} \frac{B}{H}\;, \label{Q}
 \end{eqnarray}
Replacing the above results in \rf{BE1}   we obtain the Friedman equation modified by the presence of the extrinsic curvature, i.e.,
\begin{equation}\label{eq:Friedman}
\left(\frac{\dot{a}}{a}\right)^2+\frac{\kappa}{a^2}=\frac{8}{3}\pi G\rho+\frac{b^2}{a^4}
\end{equation}
To  complete  the set  of  dynamical  equations we  use  \rf{gupta}.   Applying   \rf{k}  to  the  definition \rf{fmunu}  we  obtain  for  the  FLRW  metric
\begin{equation}
f_{ij}  =\frac{2}{K} g_{ij},\; i,j  = 1..3, \; \;\;  f_{44}  = -\frac{2}{K}\frac{1}{\dot{a}}\frac{d}{dt}{\left(\frac{b}{a}\right)}
\label{eq:fij}
\end{equation}
where  the  function $b(t)=k_{11}$  remains  undefined.  To  find  it  we   submit    it to  \rf{gupta}.   The  main  difficulty  here  is  the  determination of  the  source ${\cal T}_{\mu\nu}$  of   that  equation   for,   if  for  no  other philosophical reasons (e.g.  if  the universe  expands,  it  expands  to  where?),  we  have  no previous  experience   on  the  dynamics  of  space-time  deformations.  In this case, the   correct procedure  is  to look  for  models  that   fit  the  experimental  data  on  the  expansion of  the universe,  as  for  example  the  perfect  fluid  used     in   \cite{GDEI}.
However, within the  context  of  a    geometry   and  topology   of  the universe   determined  by the  observations,
 the   acceleration   of  the  universe    can   be   seen   as  a the observable  effect  associated  with  the   deformation of  the universe   defined  by the   extrinsic  curvature. The  simplest   option for the   external  source  of   equation \rf{gupta}   is  the  void      characterized  by    $\mathcal{T}_{\mu\nu}=0$. We   regard  this  a   a  first    attempt   to  see   where    such  reasoning  leads  us.

 The   Ricci-flat-like    equation   \rf{gupta}     becomes   simply
 \be
{\cal  F}_{\mu\nu}=0   \label{eq:guptaflat}
\ee
From \rf{fij}   we  derive  the  components of  \rf{upsilon}; of the
f-curvature  ${\cal F}_{\mu\nu\rho\sigma}$
  and finally  write  the Ricci-flat  equation   \rf{guptaflat},  whose  components   ${\cal F}_{\mu\nu}$ in this  particular    example  are
%\begin{widetext}
\begin{equation}
\mathcal{F}_{11}=\frac{1}{4} \frac{-4b^2\xi\dot{K}^2 \!+\! 5b\xi\dot{K}\dot{b}K-\dot{b}^2\xi K^2 + 2b^2\xi K \ddot{K} -
2b\ddot{b}\xi K^2 - b^2\dot{K}\dot{\xi}K + bK^2\dot{b}\dot{\xi}}{\xi^2 K^2 b} \label{eq:11}=0
\end{equation}
\begin{equation}
\mathcal{F}_{22}=r^2 \frac{-4b^2\xi\dot{K}^2 + 5b\xi\dot{K}\dot{b}K-\dot{b}^2\xi K^2 + 2b^2\xi K \ddot{K} - 2b\ddot{b}\xi K^2 - b^2\dot{K}\dot{\xi}K + bK^2\dot{b}\dot{\xi}}{4\xi^2 K^2 b}=0
\end{equation}
\begin{equation}
\hspace{-9cm}\mathcal{F}_{33}=\sin^2(\theta)\mathcal{F}_{22}=0
\end{equation}
\begin{equation}
\mathcal{F}_{44}=-\frac{3}{4} \frac{\dot{b}^2\xi K^2 + 2b^2\xi K \ddot{K} -2b\ddot{b}\xi K^2 - b^2\dot{K}\dot{\xi}K + b K^2\dot{b}\dot{\xi} -2b^2\xi \dot{K}^2 + b \xi K \dot{K}\dot{b}}{\xi K^2  b^2} =0 \label{eq:44}
\end{equation}
%\end{widetext}
 The  only essential equations in the  above  set  are  the  first  and  last  equations. By subtracting these equations we obtain $b^2\dot{K}^2 + K^2\dot{b}^2 = 2bK\dot{b}\dot{K}$ or, equivalently,
\be
\left(\frac{\dot{K}}{K}\right)^2 - 2\frac{\dot{b}}{b} \frac{\dot{K}}{K} = -\left(\frac{\dot{b}}{b}\right)^2
\ee
which  has  a   simple  solution   $K(t)= 2\eta_0 b(t)$,  where  we   have   denoted  by $2\eta_0$ its  integration  constant.
Replacing the  expression of  $K$  given   by  (\ref{eq:hk}), we obtain
\begin{equation}
\label{eq:EG1}
\frac{B}{H}=1\pm \sqrt{4 \eta_0^2 a^4 - 3}
\end{equation}
Of  course,  to  obtain real  values of  $a$  and  $b$,
we  must have  the   condition
\be
\eta_0^2  \ge  \frac{3}{4}\frac{1}{a^4}  \label{eq:eta}
\ee
Expressing   $Q_{\mu\nu}$   in  terms  of  $B/H$  given  by  \rf{Qab},  the  conservation  equation \rf{cons} can  be
readily  integrated   giving
\be
2\frac{B}{H} -1  =\beta_0  \label{eq:consQ}
\ee
where  $\beta_0$  is  a   second  integration  constant.

Subtracting  \rf{consQ} from    \rf{EG1},  we obtain
the searched    equation on $b(t)$ expressed  as a function of the expansion parameter $a(t)$
\be
\frac{\dot{b}}{b} =\frac{\dot{a}}{a}( \beta_0  \mp \sqrt{4\eta_0^2 a^4 -3})
\ee
The integration of  which  is   very  simple.  Merging all
  integration constants  into a  single one  $\alpha_0$
the  final  solution  can be expressed  as
\begin{equation}
b(t)  =  \alpha_0 a^{\beta_0} e^{\mp \gamma (a)}  \label{eq:b1}
\end{equation}
where  $\gamma(a)$ is  given  by
\begin{equation}
\!\!\gamma(a) = %\int \frac{1}{a}\sqrt{4\eta_0^2 a^4 - 3}\; da \!\!=
\! \!\sqrt{4\eta_0^2 a^4 - 3}- \! \sqrt{3}\arctan\left(\frac{\sqrt{3}}{3}\sqrt{4\eta_0^2 a^4 - 3}\right)
\label{eq:gamma}
\end{equation}
Replacing  \rf{eta}  and  \rf{b1}   in  \rf{Friedman} we  obtain the  Friedman  equation modified  by   the  extrinsic  curvature:
\begin{equation}
\label{eq:FriedmanNash}
\left(\frac{\dot{a}}{a}\right)^2+\frac{\kappa}{a^2}=\frac{4}{3}\pi G\rho+\frac{\alpha_0^2 a^{2\beta_0} e^{\mp2 \gamma (a)}}{a^4}
\end{equation}
As we see   the result  depends  on   a  choice of   three  integration constants  $\alpha_0$,   $\beta_0$ and   $\eta_0$
which  must be   adjusted   by  known  boundary   conditions:

(a) The  constant    $\alpha_0$   is  a  scale  factor  for  $b(t)$
which  can  be    fixed  once  for  all  for  today's  ($t=0$) value to be   $\alpha_0=1$,   by  setting $b(0)=a^\beta_{0}(0) e^{\mp \gamma(a=1)}$,  where we  have  denoted   today's  value  of   $$e^{\gamma \,(a=1)}=\sqrt{4\eta_0^2 - 3}- \sqrt{3}\arctan\left(\frac{\sqrt{3}}{3}\sqrt{4\eta_0^2 - 3}\right)$$

(b) The equal sign in  \rf{eta} gives   $\gamma (a)=0$,  which   corresponds  to  the particular  case   previously  studied  in   our  previous  paper \cite{GDEI}, where  a  comparison  of  the  extrinsic  curvature   with  a phenomenological  fluid  (the  X-fluid)   was used.
In the   following   we  consider   the   more general  cases   corresponding  to the greater sign  ($>$) in  \rf{eta}.

\vspace{2mm}

In order  to  evaluate  the  above  results  with  the  presently   available  data  we  translate  the   equations   in  terms of  the  redshift  $z$,   when    the   expansion  parameter becomes  $a(z) = 1/(1+z)$ and the  condition  \rf{eta}  becomes  $\eta_0^2 \ge \frac{3}{4}(1+z)^4  $.
Furthermore,  we  express  \rf{FriedmanNash}  in terms  of  the  relative  densities $\Omega_k,\;\Omega_\Lambda, \; \Omega_{matter}=\Omega_m,\; \Omega_{extrinsic}=\Omega_{ext}$,  with  the  following observations.

(1) Since the   value  of  the  spatial  curvature  $\kappa$ in  \rf{FriedmanNash} has been consistently   verified  to be  zero \cite{wmap}, we  will  simple  ignore  the  contribution  of $\Omega_k$.

(2) From our  previous  arguments  on the
cosmological  constant problem   we  have  eliminated  the cosmological  constant   contribution  in  this  analysis,  so that we  also take $\Omega_{\Lambda}=0$. We   will  see that  the  contribution  of  $\Lambda$  is not really relevant to the  accelerated  expansion   in  presence of  the contribution of the extrinsic curvature.

(3) The   baryonic     matter  relative  density is  denoted  by   $\Omega_m$  and  the  extrinsic  relative  density is  denoted  by   $\Omega_{\rm ext}$.  Assuming  the  standard  normalization condition  $H\rfloor_{z=0} =H_0=100h \; km. s^{-1}\;Mpc^{-1}$  (the  Hubble  constant), we may  write these  in terms  of  $z$  as
\be
\Omega_m =\frac{8\pi G}{3 \rho (1+z)^3}\;\;  \mbox{and} \;\; \Omega_{\rm{ext}} =  \frac{1-\Omega_m}{e^{\gamma \,(z=0)}} \label{eq:omegaext}
\ee

With  these   considerations   the  modified  Friedman   equation  \rf{FriedmanNash}   written  in terms  of the redshift becomes
%\begin{widetext}
\begin{equation}
 \label{eq:fr}
{{E(z)}} =  \frac{\dot{a}(z)}{a(z)}=
\left[\Omega_m(1+z)^{3}  + {\Omega_{\rm{ext}} (1+z)^{4-2\beta_0}}\right]^{1/2}
\end{equation}
%\end{widetext}
\vspace{2mm}\\

To  find  if  this   result  corresponds   to the  observations
we  use a  statistical  analysis which  gives  a    model independent  probe of the  accelerating expansion of the universe \cite{wmap}.  This  is given  by   the  dimensionless \emph{luminosity-distance}  expression
\begin{equation}
d_L(z)= (1+z)\frac{\int_0^z \frac{dz'}{E(z')}}{ H_0}
\end{equation}

For  the   two  considered  density parameters  $\Omega_m$ and $\Omega_{\rm{ext}} $,  the  luminosity  distance  is  related  to the \emph{distance  modulus}  (with $d_L(z)$   measured  in  Mpc) as
\[
\mu(z,u) =m-M =5\, log\, d_{L(z)}  +  25
\]
where  the     parameters     $m$   and   $M$ represent   respectively  the   apparent  and  absolute bolometric
magnitudes  \cite{Lixin}.

We  may  evaluate  the   contribution of   the  extrinsic  curvature  by   plotting  the   contours  in the planes
$(\Omega_m, \beta_0)$  for  different  values  of $\eta_0$.

For   the  SN Ia  database, the  best   fit  values is   given  by
the   likelihood   analysis is  based on the  calculation of
the   standard  distribution
\[\chi^2(u)= \sum^{115}_{i=1} \frac{\left[\mu^i_{p}(z|\mathbf{u}) - \mu^i_{0}(z|\mathbf{u}) \right]^2}{\sigma^2_i}
\]
where $\mu^i_{0}(z|\mathbf{u})$ is the extinction
corrected distance modulus for a given SNe Ia at $z_i$ and $\sigma_i$ is the standard deviation of the uncertainty in the individual distance moduli  (including uncertatinties in  galaxy  red shifts).  The above  summation   was  taken   over the 115 observational Hubble data for  SN Ia  at    redshifts $z_i$
 \cite{snls} (For more  details on such    SN Ia   statistical  analysis  we refer the reader to~\cite{sne,sn1,sn2,sn3,sn4,sn5,sn6} and refs. therein.). We  may  estimate  the  admissible values  of    $\beta_0$  for  the    best  fit  values   of    the known  data set on  SN Ia in the parametric plane  $(\Omega_m, \beta_0) $, with  constant   $\Delta \chi^2 = 2.30,\;\; 6.17, \;\; 11.8$,  respectively  for   $\eta_0 = 3.5, 5.0, 7.0$,  corresponding  to  the above mentioned 115 observations. The  first value $\eta_0 > 3.5$  was  taken  from  \rf{eta}.  The other two values, i.e., $\eta_0 = 5.0$ and $\eta_0 = 7.0$ were taken arbitrarily  in  the sequence.

Using  data  from   \cite{snls} and  since the highest-$z$ supernova  Ia in our sample is at $z \simeq 1.01$ at 68.3\% (C.L.) we have found  for  the  three  above  values  of  for $\eta_0$,  respectively
\[
\mbox{for}\;\;\eta_0= 3.5, \quad \beta_0 = -1.45^{+0.30}_{-0.25} \quad \mbox{and} \quad \Omega_m = 0.14 \pm 0.03\;,
\]
\[
\mbox{for}\;\; \eta_0=5.0,\quad \beta_0 = -3.09^{+0.5}_{-0.4} \quad \mbox{and} \quad \Omega_m = 0.20 \pm 0.03\;,
\]
\[
\mbox{for}\;\; \eta_0=7.0, \quad \beta_0 = -5.35^{+0.7}_{-0.6} \quad \mbox{and} \quad \Omega_m = 0.24 \pm 0.03\;.
\]
By combining the above results with the normalized  expression  in
\rf{omegaext},  we  may  estimate that the extrinsic curvature density parameter lies   in the interval
 $$10^{-2} \ge \Omega_{\rm{ext}} \ge 10^{-6}$$
showing  that   there  is  a    wide  range  of  the parameters  for  the    extrinsic  curvature  density  which   fit  the  observations.

\begin{figure}[h!]
\centerline{
\psfig{figure=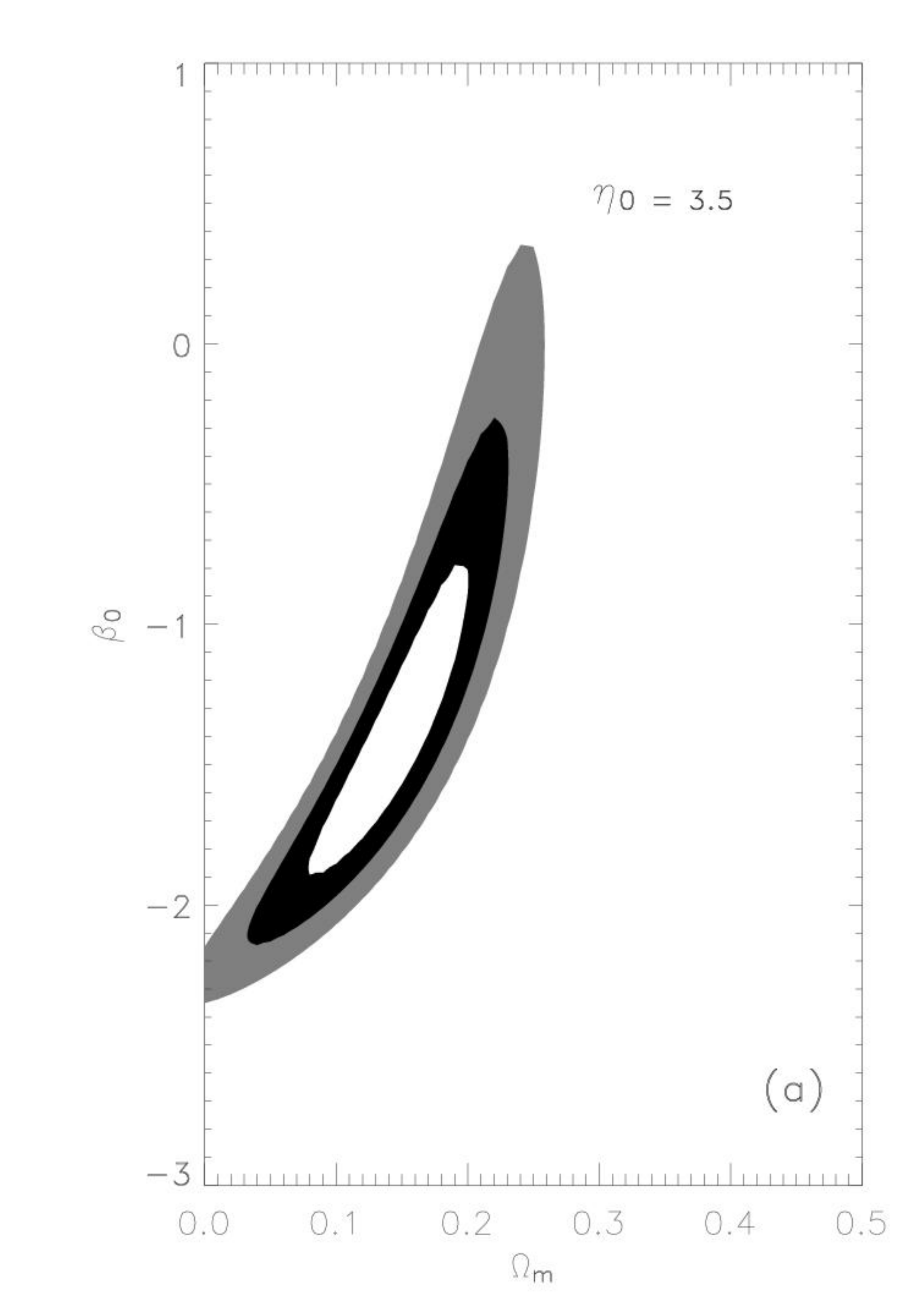,width=2.3truein,height=2.7truein,angle=0}
\psfig{figure=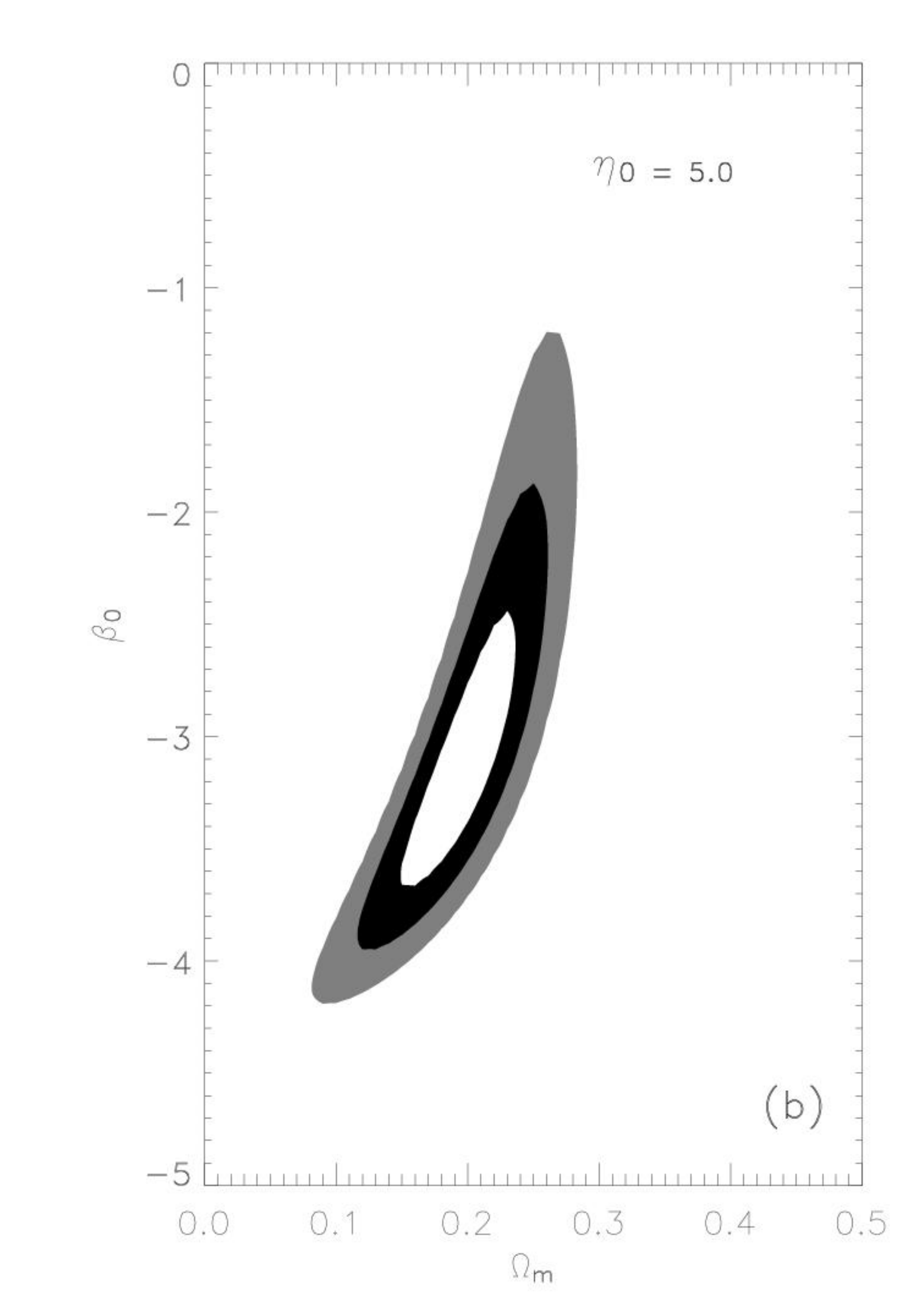,width=2.3truein,height=2.7truein,angle=0}
\psfig{figure=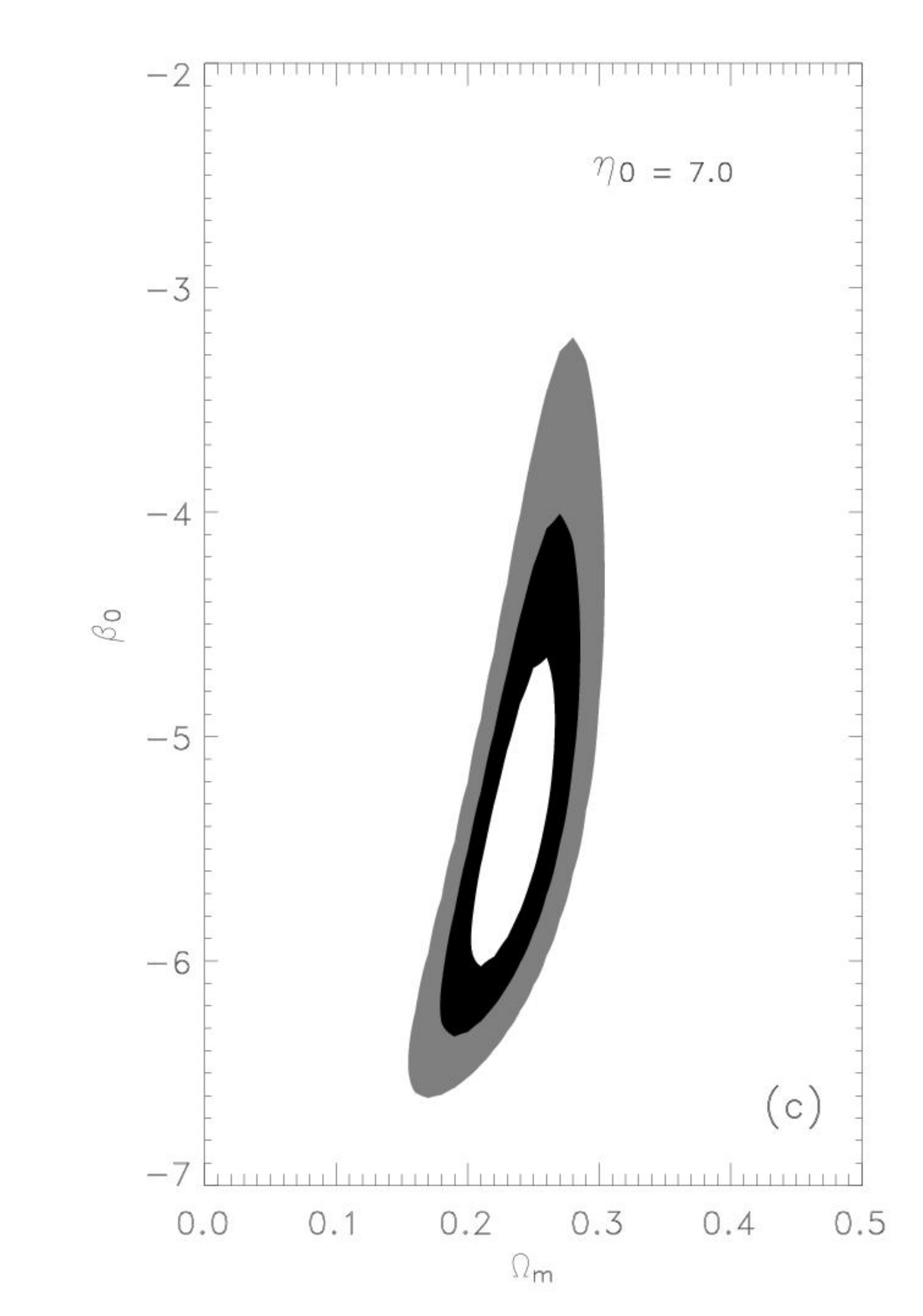,width=2.3truein,height=2.7truein,angle=0}
}
\caption{Contours of the   $\chi^2$ test  in the parametric space $\Omega_m$ (horizontal  axis)  versus  $\beta_0$ (vertical axis). The contours are drawn for $\Delta \chi^2 = 2.30$,  $6.17$  and $11.8$. As explained in the text, the value of $\eta_0$ has been fixed at 3.5 (a), 5.0 (b) and 7.0 (c). In particular, we note that for $\eta_0 = 7.0$, the allowed   $\sigma$ interval for the matter density parameter is very close to that provided by current dynamical estimates, i.e., $\Omega_m \simeq 0.2-0.3$.}
\end{figure}

As  a  last  remark  we  note  that    the  contribution  of  the
extrinsic  curvature  is also  consistent  with the  expected age   of  the  universe.  This  can  be  seen
directly   from  \rf{fr},  from  which  we  extract  the
the  age  of  the  universe
\[
t  = \frac{-1}{H_0}  \int{\frac{dz}{  (1+z)\sqrt{\Omega_m (1+z)^3  +\Omega_{\rm ext}(1+z)}}}
\]
From  this  expression  we  conclude  from  the  contour  (b)  in  the  above  figure  that  for  $0.14 \le \Omega_m \le 0.3$,
we  obtain  the  age  of  the  universe  between  $12\le  t  \le 16$,  which is  compatible  with  the  estimated formation  of  the  large  structures  \cite{Komatsu}.

\begin{center}
$\textbf{Summary}$
\end{center}
We have applied   the  concept  of    smoothly     deformable
Riemannian  manifolds  to  relativistic  cosmology.
The  concept   is  similar  to  the  one  used by  Perelman's  solution  of  the Poincaré\'e  conjecture,   but  where  we  applied  Nash's  deformation  instead  of  the  Ricci flow.  %
The  advantage  Nash's  geometric  flow  condition  over the Ricci  flow  is  that  it is    entirely relativistic  and  compatible  with   Einstein's  equations.  However,  Nash's   geometric  description  involve  a  new  variable, the  extrinsic  curvature,    so  that  it  also requires  a  new dynamical process.

With  basis   in the    spin-statistic  theorem  we   have  suggested  an   Einstein-like   dynamical  equation  for  the   extrinsic  curvature adapted  from  the  original  equation  of  S. Gupta. The  result  for  a massless  spin-2  field
show  that  the when  the   deformation of the  geometry  produced  by  the   extrinsic  geometry  is   applied  to  the universe,   we  obtain  a  consistency   with  the  current observations.  We  have  applied  a   model independent statistical  analysis,  showing   that  the   cosmological  constant   does not  play  a  significant  role  on the  acceleration of the  universe  in  presence  of  the  deformation,  at  least  within the present  observational  range.

The  deformation process defined by  Nash  requires  the  embedding  of  the space-time  in  a  larger  space. However, since the  standard   gauge  fields  which   are  required  for  our experimental basis  are  defined only in  four-dimensions,  the  end  result  is   a   four-dimensional  deformed  space-time  which is
obtained  by  the   inverse  embedding  map.  The   four-dimensional  observers  with its  gauge field  based  technology   will  measure  the  end  effects  of  the  deformations  without  being  aware   of  the  embedding.

The  presence  of  the extrinsic   curvature  leads also to  a  new   conserved  quantity the  deformation  tensor  $Q_{\mu\nu}$,  and   so   to  an  observational  effect  which   adds    some  topological  qualities to  Einstein's  gravitation  theory.  This  interpretation  is  supported  by   the  Gauss  and  Riemann views  that  the true   geometry   will  at the  end  be  determined    by  the  observations.  Our   estimates    suggest that the observed  acceleration   of  the  universe   evidences  the  existence  of   a  deformation  at the  cosmological  scale, giving  to  the  universe   some  notion of  its    shape.

\end{document}